\documentclass[useAMS,usenatbib]{mn2e}

\usepackage{graphicx}
\usepackage{amsmath}
\usepackage{amssymb}

\title[Evidence for a warm ISM in Fornax dEs]{Evidence for a warm ISM
  in Fornax dEs -- II. FCC032, FCC206 and FCCB729}
\author[D. Michielsen, S. De Rijcke, W. W. Zeilinger, P. Prugniel,
  H. Dejonghe, S. Roberts]{D. Michielsen$^{1}$\thanks{E-mail:
  dolf.michielsen@ugent.be}, S. De Rijcke$^{1}$\thanks{Postdoctoral
  Fellow of the Fund for Scientific Research - Flanders, Belgium (F.W.O.)},
  W.~W. Zeilinger$^{2}$, P. Prugniel$^{3,4}$, H. Dejonghe$^1$, S. Roberts$^{5}$\\
  $^{1}$Sterrenkundig Observatorium, Ghent University, Krijgslaan 281, S9, B-9000 Gent, Belgium \\
  $^{2}$Institut f\"ur Astronomie, Universit\"at Wien, T\"urkenschanzstra{\ss}e 17, A-1180 Wien, Austria\\
  $^{3}$CRAL-Observatoire de Lyon, CNRS UMR 142, F-69561 St-Genis-Laval Cedex, France\\
  $^{4}$GEPI, Observatoire de Paris Meudon, F-92190 Meudon, France\\
  $^{5}$School of Physics and Astronomy, Cardiff University, PO Box 913, Cardiff, CF24 3YB, UK
}

\begin{document}
	 
\date{}
\pagerange{\pageref{firstpage}--\pageref{lastpage}} \pubyear{2004}

\maketitle \label{firstpage}

\begin{abstract}
  We present $R$-band and H$\alpha$+[N\,\textsc{ii}] narrow-band
  imaging of FCC032, FCC206 and FCCB729, three dwarf elliptical
  galaxies (dEs) in the Fornax Cluster. These dEs contain significant
  amounts of ionized gas. FCC032 harbours a large ionized gas complex,
  consisting of several individual clouds, a superbubble and a
  filament that extends away from the galaxy centre. The ionized gas
  structures observed in FCC032 bear a strong resemblance to those
  observed in more gas-rich and more fiercely star-forming dwarf
  galaxies. FCC206, a very low surface brightness dE, contains one
  faint extended emission region, and two compact clouds. In FCCB729,
  the only nucleated galaxy in this sample, one of the ionized gas
  clouds coincides with the stellar nucleus. We derive ionized gas
  masses of a few $10^2$ to $10^3$\,M$_\odot$ for these galaxies.\\
  This brings our sample of dEs with ionized gas with
  H$\alpha$+[N\,\textsc{ii}] narrow-band imaging to five objects. The
  ionized gas morphologies in these galaxies range from pure nuclear
  emission peaks to extended emission complexes. This morphological
  diversity could also indicate a diversity in ionising processes in
  dEs with a warm interstellar medium (active galactic nuclei,
  starbursts, post-asymptotic giant branch stars,\ldots).\\
  Inside FCC206, four star clusters appear to be on the verge of
  merging to form a nucleus in this presently non-nucleated
  dE. Understanding the formation of nuclei in dEs could give us
  important clues to the formation of super-massive black holes (SMBHs).
\end{abstract}


\section{Introduction} \label{sec_intro}

Dwarf elliptical galaxies (dEs) are the most abundant galaxy type in
clusters and groups of galaxies. They are faint, with exponentially
declining surface brightness profiles \citep{fb94}. Until recently,
dEs were thought to have lost their gas and dust long ago. Several
evolutionary scenarios have been proposed to explain this apparent
lack of an interstellar medium (ISM) in dEs. The 'wind-model'
(e.g. \citet{mori97}) proposes that dEs are primordial objects that
lost their gas after it was heated above the escape velocity by
supernova explosions.  Alternatively, the frequent high-speed
interactions with giant cluster-members to which a small late-type
disc galaxy (Sc-Sd) is subjected and the subsequent starbursts can
transform it into a gasless spheroidal dE-like object. This `galaxy
harassment' process \citep{moore96} induces a dramatic morphological
evolution on a time-span of about 3\,Gyr.  Moreover, hydrodynamical
simulations of dwarf galaxies moving through the hot, thin
intergalactic medium in clusters \citep{mb00} or groups
\citep{marcollini03} show that ram-pressure stripping can completely
remove the ISM of a dwarf galaxy less massive than $10^9$\,M$_\odot$
within a few 100\,Myr. If the ram pressure is strong enough, the
interstellar medium may be compressed and localised star bursts may
ensue. Low-mass dwarfs ($M \sim 10^8$\,M$_\odot$) lose their gas very
rapidly, leaving virtually no time for star-formation after entering
the cluster of group. Yet another possibility is that dEs are related
to other dwarf galaxies such as Blue Compact Dwarfs (BCDs). The
`fading model' conjectures that star-forming dwarf galaxies will fade
and reach an end-state similar to present-day dEs after they have used
up their gas supply and star-formation has ended, although some
controversy still remains \citep{drink91,marlowe99}. Interactions may
have sped up the gas-depletion process, explaining both the abundance
of dEs and the paucity of BCDs in high-density environments. These
different scenarios to explain the origin of dEs are not mutually
excluding. An important common point between the propositions is that
they all remove the gas from dwarf galaxies. In addition, unlike
massive galaxies, dEs can hardly acquire gas in collisions since their
escape velocity is small compared to the velocity dispersion in galaxy
clusters.

For all these reasons, dEs were thought to be gas-depleted
systems. However, a growing amount of evidence indicates that at least
some dEs have retained part of their gas. Since the efficiency of
galaxy interactions and ram pressure at stripping the gas off a dwarf
galaxy obviously depends on the dwarf's orbit, dEs on high angular
momentum orbits in the outskirts of a cluster can be expected to be
more gas-rich.  This has indeed been observed in an H\,\textsc{i}
study of a sample of Virgo dEs by \citet{conselice03}. They find that
H\,\textsc{i} detected dEs are preferentially located near the
periphery of the Virgo Cluster, with an overall detection rate of
about 15 per cent.  In a spectroscopic survey of the Fornax Cluster,
\citet{drink01} discovered H$\alpha$ emission in about 25 per cent of
the dEs. Most of these galaxies also lie towards the outskirts of the
cluster, while dEs near the centre of the cluster are generally devoid
of ionized gas, again suggesting an environmental effect on the
gas-depletion rate of dEs (e.g. interactions or ram-pressure
stripping) and strongly hinting at an (evolutionary) transition
between gas-rich/star-forming dEs at the outskirts and
gasless/quiescent ones in the centre of the cluster.

This paper is the second in a series on H$\alpha$ imaging of the
ionized ISM of Fornax dEs. In \citet{derijcke03b} (hereafter Paper~1)
results on FCC046 and FCC207 are reported (for FCC046, see also De
Rijcke \& Debattista 2004). Both dEs exhibit a central emission region
that could be attributed to photo-ionization by post-asymptotic giant
branch (post-AGB) stars. FCC046 also harbours 6 faint emission clouds
with diameters and fluxes comparable to supernova remnants. The
hypothesis that FCC046 is actively forming stars, albeit at a
leisurely pace, is further corroborated by its blue colour and high
near-infrared Paschen absorption index \citep{michielsen03}. Such dEs
could be considered as the missing link between more vigorously
star-forming dwarfs (such as BCDS) and traditional dEs.

Here, we present $R$-band and H$\alpha+$[N\,\textsc{ii}] narrow-band
images of three Fornax dEs (FCC032, FCC206, and FCCB729) that are
expected to contain ionized gas, based on the existing spectroscopic
equivalent-width (EW) estimates \citep{drink01}. In section
\ref{sec_obs}, we discuss the details of the observations and data
reduction. The results based on the $R$-band and the
H$\alpha+$[N\,\textsc{ii}] narrow-band images are presented in
sections \ref{sec_rband} and \ref{sec_halpha}, respectively, and
discussed in section \ref{sec_disc}. We summarise our conclusions in
section \ref{sec_concl}.  Throughout the paper, we use
$H_0=75$\,km\,s$^{-1}$\,Mpc$^{-1}$ and a Fornax systemic velocity
$v_\text{sys} = 1379$\,km\,s$^{-1}$, which gives a distance of
18.4\,Mpc to the Fornax Cluster..

\section{Observations and data reduction}  \label{sec_obs}

The observations were carried out on 2003 October 18 and 19 with Yepun
(VLT-UT4) using FORS2 in service mode. We took 16-min exposures with
the H\_Alpha/2500+60 filter centred on 6604\,{\AA} (this redshifted
H$\alpha$ filter gives the best overlap with the H$\alpha$ emission
line in galaxies at the redshift of Fornax) and with a
FWHM\,$=64$\,{\AA}. 130--160\,s exposures with the R\_Special$-$71
filter centred on 6550\,{\AA} and with a FWHM\,$=1650$\,{\AA} were
taken to serve as off-band images. H$\alpha$ images of the
spectrophotometric standard stars LB227 and LTT2415 were taken for
flux-calibration. During the observations, the seeing typically was
0.5 -- 0.7\,arcsec FWHM (determined from the stars on the images). The
standard data reduction procedures (bias subtraction, flatfielding,
cosmic removal, sky subtraction) were performed with
MIDAS\footnote{ESO-MIDAS (Munich Image Data Analysis System) is
developed and maintained by the European Southern Observatory}. Before
co-adding, all science images were corrected for atmospheric
extinction, using the $R$-band extinction coefficient : $k_c = 0.068$
provided by the ESO Quality Control, and interstellar extinction,
using the Galactic extinction estimates from \citet{schlegel98} : $A_R
= 0.034$ for FCC032, $A_R = 0.038$ for FCC206 and $A_R = 0.059$ for
FCCB729. The images were finally converted to units of electron
s$^{-1}$ pixel$^{-1}$.

In order to find the correct scaling for the $R$-band images we adopted
the following strategy. The pure emission `Em' can be recovered from a
narrowband image `Ha' and an $R$-band image `Rb' as
\begin{equation}
  \text{Em} = \text{Ha} - (c \times \text{Rb} - \delta),
  \label{eq_Em}
\end{equation}
with $c$ the proper scaling constant and $\delta$ a correction for
possible faulty sky-subtraction. To find the best values for $c$ and
$\delta$, we fitted the isophotes of the narrow-band and $R$-band
images in an annulus between $m_R = 24.5$ and $m_R =
26.5$\,mag\,arcsec$^{-2}$ which in retrospect did not contain any
emission (hence Em\,$= 0$), using the standard MIDAS FIT/ELL3
command. Thus, a smooth version of this annulus could be constructed
for both images. The optimal values of $c$ and $\delta$ can be found
by minimizing the expression $|\text{Ha} - (c\times\text{Rb}+\delta)|$
with Ha and Rb the smoothed versions of the annulus. With these values
in hand, the pure-emission image can be obtained using relation
\eqref{eq_Em}.

For all galaxies, $\delta$ was essentially zero, which makes us
confident that the sky was properly subtracted from all images. Since
the H$\alpha$ narrowband and $R$-band filters overlap, subtracting an
$R$-band image instead of a narrowband continuum image induces a
partial removal of some H$\alpha$+N[\textsc{ii}] light. The error thus
introduced is of the order of the ratio of the effective widths of the
filters ($R$-band:\,$W=1650$\,{\AA} and
H$\alpha$:\,$W=64$\,{\AA}). This is less than 4 per cent which is
negligible in comparison to other possible sources of error, so we did
not correct for it.

A pixel-value (in electrons s$^{-1}$) in the
pure-emission image, denoted as $N$, can be converted to flux units
(W\,m$^{-2}$), denoted by $F^\prime$, using the formula
\begin{equation}
  F^\prime = N \times \frac{\varphi_0}{N_*}\int_0^\infty {\cal
    F}_*(\lambda) \varphi_f(\lambda) d\lambda \text{\,W\,m}^{-2}.
\end{equation}
Here, ${\cal F}_*(\lambda)$ is the spectrum of a flux-calibration
standard star in W\,m$^{-2}$\,{\AA}$^{-1}$ and $N_*$ is the measured
flux of that star, expressed in electron s$^{-1}$. The function
$\varphi_f(\lambda)$ is the transmission of the H$\alpha$ filter and
$\varphi_0$ the transmission of the optics (which is basically
constant for a narrow-band filter). The prime on $F^\prime$ indicates
that this is the flux incident on the CCD, after going through the
telescope and instrument optics and the narrow-band filter. This can
also be written as
\begin{equation}
  \begin{array}{lcl}
  F^\prime &=& \varphi_0 \left[
    F_{\text{H}\alpha} \varphi_f \left(\lambda_{\text{H}\alpha}\right) + 
    F_{\text{[N\,\textsc{ii}]}_1} \varphi_f \left(\lambda_{\text{[N\,\textsc{ii}]}_1}\right) \right. \\
  & & \left. \;\;\;\;\; + 
    F_{\text{[N\,\textsc{ii}]}_2} \varphi_f \left(\lambda_{\text{[N\,\textsc{ii}]}_2}\right)
    \right],
  \end{array}
\end{equation}
with $F_{\text{H}\alpha}$, $F_{\text{[N\,\textsc{ii}]}_1}$ and
$F_{\text{[N\,\textsc{ii}]}_2}$ the incoming fluxes, before going
through the telescope optics and the narrow-band filter, of
respectively the H$\alpha$ 6564{\AA}, the [N\,\textsc{ii}] 6548{\AA}
and the [N\,\textsc{ii}] 6584{\AA} (redshifted) emission lines
approximated as $\delta$-functions. This allows one to obtain the true
incoming flux of the H$\alpha$ emission line as
\begin{equation}
  F_{ \text{H}\alpha } = \frac{ 
    \frac{N}{N_*} 
    \int_0^\infty {\cal F}_* (\lambda) \varphi_f (\lambda)\, d\lambda 
  } 
  { 
    \varphi_f \left(\lambda_{\text{H}\alpha}\right) + 
    \left[ 
      \frac{ F_{\text{[N\,\textsc{ii}]}_1}}{ F_{\text{[N\,\textsc{ii}]}_2} } 
      \varphi_f \left(\lambda_{\text{[N\,\textsc{ii}]}_1}\right) + 
      \varphi_f \left(\lambda_{\text{[N\,\textsc{ii}]}_2}\right) 
    \right]
    \frac{F_{\text{[N\,\textsc{ii}]}_2} }{ F_{\text{H}\alpha} } 
  }.
  \label{eq_fluxha}
\end{equation}
The total incoming H$\alpha$+[N\,\textsc{ii}] flux is simply
\begin{equation}
  F_{\text{em}} = F_{\text{H}\alpha}
  \left( 1 +
  \frac{ F_{\text{[N\,\textsc{ii}]}_1} }{ F_{\text{H}\alpha} } + 
  \frac{ F_{\text{[N\,\textsc{ii}]}_2} }{ F_{\text{H}\alpha} }
  \right) .
  \label{eq_fluxem}
\end{equation}
Since the H$\alpha$ filter is relatively flat-topped and the
[N\,\textsc{ii}] lines are well inside the filter transmission curve,
the total flux is rather insensitive to the adopted relative
line-strengths. In the following, we will assume the mean value
$F_{\text{[N\,\textsc{ii}]}_2}/F_{\text{[N\,\textsc{ii}]}_1} = 3$ for
the ratio of the line-strengths of the two N lines
\citep{phillips86}. The ratio
$F_{\text{[N\,\textsc{ii}]}_2}/F_{\text{H}\alpha}$ is not known and is
treated as a free parameter, though it is generally assumed that it
varies between 0 and 2.

\begin{figure}
  \includegraphics[width=84mm]{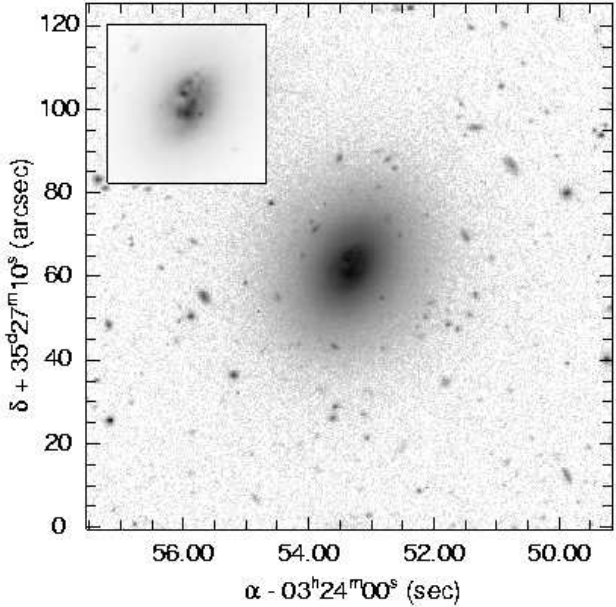}
  \caption{130 sec. negative $R$-band image of FCC032 in a logarithmic
    grayscale between 30 and 21.5 mag. The inset shows the inner $25
    \times 25$\,arcsec with a linear grayscale. Note the many bright
    knots (black) interspersed with dust patches (see also
    Fig. \ref{fig_032un}).}
  \label{fig_032rb}
\end{figure}
\begin{figure}
  \includegraphics[width=84mm]{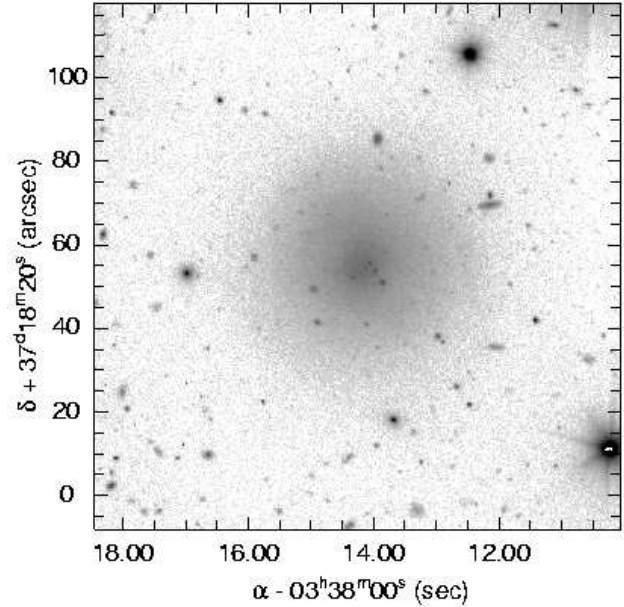}
  \caption{160 sec. negative $R$-band image of FCC206. The adopted
    grayscale is the same as in Fig. \ref{fig_032rb}, allowing a
    direct comparison of the surface brightness. Note the four point
    sources close to the galaxy centre (see also Fig. \ref{fig_206un}).}
  \label{fig_206rb}
\end{figure}
\begin{figure}
  \includegraphics[width=84mm]{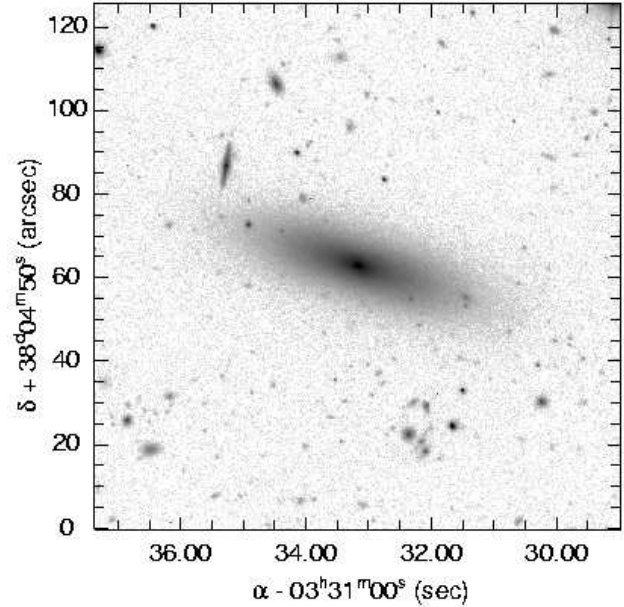}
  \caption{130 sec. negative $R$-band image of FCCB729. The adopted
    grayscale is the same as in Fig. \ref{fig_032rb}, allowing a
    direct comparison of the surface brightness.}
  \label{fig_729rb}
\end{figure}

\section{$\bmath{R}$-band photometry}  \label{sec_rband}

\begin{figure*}
  \includegraphics[clip,scale=0.9]{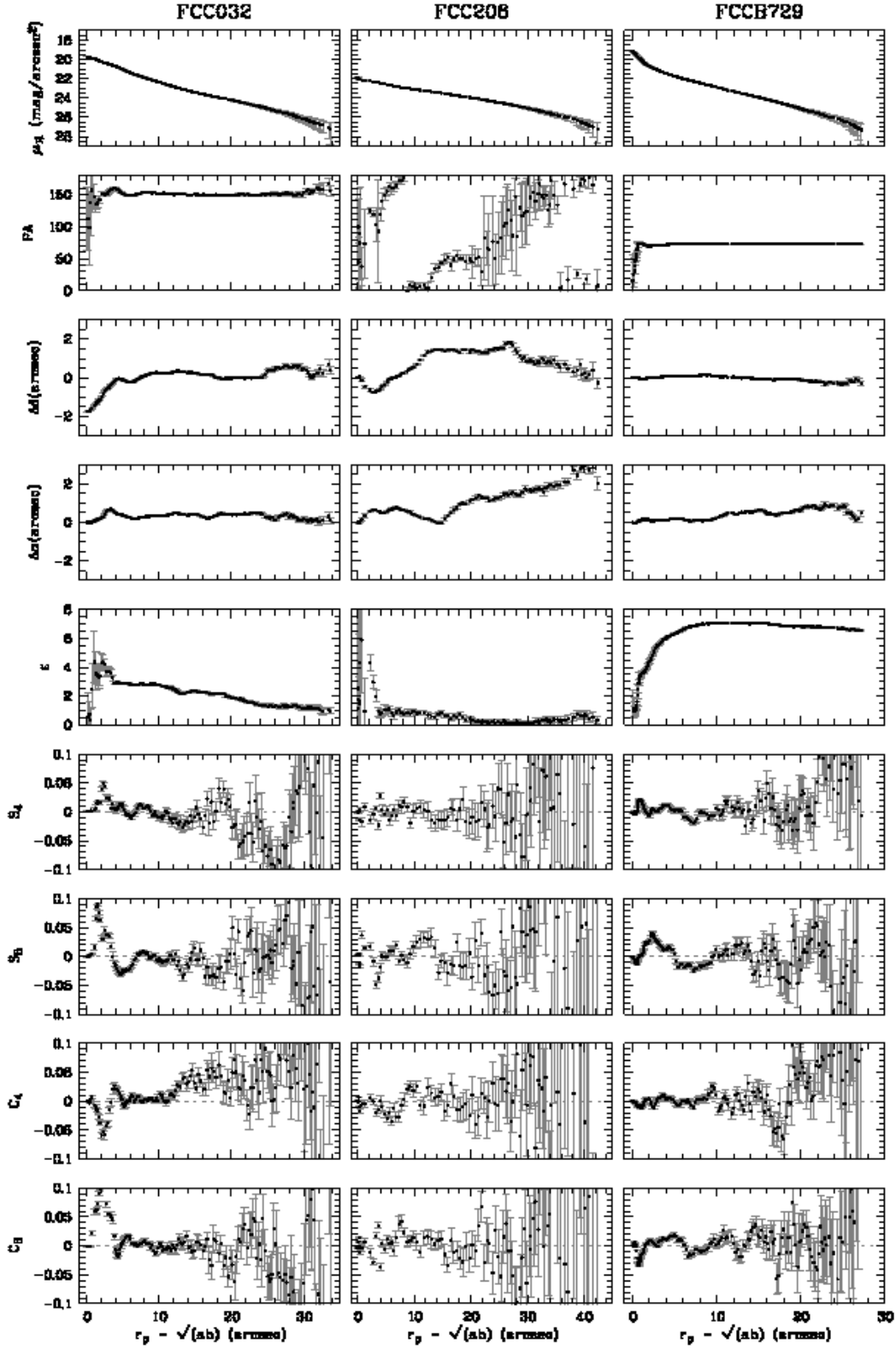}
  \caption{Photometric properties of FCC032 (left column), FCC206
    (middle), and FCCB729 (right), derived from the $R$-band images,
    versus the geometric mean of the semi-major and semi-minor
    distances $a$ and $b$. From top to bottom:~the $R$-band surface
    brightness $\mu_R$, the deviation in declination $\Delta \delta$
    and right ascension $\Delta \alpha$ of the centres of the
    isophotes with respect to the brightest point, the PA, the
    ellipticity $\varepsilon = 10\,(1- b/a)$ and the Fourier
    coefficients $S_4, S_3, C_4$ and $C_3$ that quantify the
    deviations of the isophotes from ellipses.}
  \label{fig_photometry}
\end{figure*}

We fitted ellipses to the isophotes of $R$-band
images of FCC032, FCC206, and FCCB729 (see Figs. \ref{fig_032rb},
\ref{fig_206rb}, and \ref{fig_729rb}), with the semi-major axis $a$,
the position angle (PA), the ellipticity $\epsilon = 10(1-b/a)$ and
the coordinates of the centre of each ellipse as free parameters. We
masked hot pixels, stars, and cosmics. Masked regions were not used in
the fit. For the innermost isophote, the ellipse was centred on the
brightest pixel, excluding the bright point sources in FCC032 and
FCC206. The deviations of the isophotes from a pure elliptic shape
were quantified by expanding the intensity variation along an
isophotal ellipse in a fourth order Fourier series with coefficients
$S_4, S_3, C_4$ and $C_3$:
\begin{equation}
  \begin{array}{lcl}
    I(\theta) &=& I_0 \left( 1 
      + C_3 \cos(3(\theta - \text{PA})) + C_4 \cos(4(\theta - \text{PA})) 
      \right. \\
    && \left. 
      + S_3 \sin(3(\theta - \text{PA})) + S_4 \sin(4(\theta - \text{PA})) 
      \right),
  \end{array} \label{eq_c4}
\end{equation}
with $I(\theta)$ the intensity on the ellipse at an angle $\theta$
with respect to the major axis and $I_0$ the average intensity on the
ellipse. The photometry of all three galaxies is presented in
Fig. \ref{fig_photometry}. All photometric parameters were fitted by
cubic splines as functions of the semi-major axis $a$ of the isophotal
ellipses. This allows us to reconstruct the surface brightness at any
given point on the sky and hence to reconstruct the $R$-band image
from the fitted photometric parameters. From this reconstructed image,
which is obviously free of cosmics and foreground stars, we derived
the total apparent $R$-band magnitude $m_R$ and the model-free
photometric parameters $r_\text{eff}$ and $\langle \mu
\rangle_\text{eff}$. The effective radius $r_\text{eff}$ is the
half-light radius of the galaxy; the effective surface brightness
$\langle \mu \rangle_\text{eff}$ is the mean surface brightness within
$r_\text{eff}$ and can be derived from $m_R$ and $r_\text{eff}$ using
\begin{equation}
  \langle \mu \rangle_\text{eff} = m_R + 2.5\log(2\pi r_\text{eff}^2).
\end{equation}

The surface brightness profiles of dEs can be fitted quite well by a
S\'ersic profile \citep{sersic}, which is a generalisation of de
Vaucouleurs' $r^{1/4}$ and exponential laws. Several photometric
studies have used this profile to analyse dEs (see
e.g. \citet{ryden99}, \citet{barazza03}). We fitted the surface
brightness profiles with
\begin{equation}
  \mu(r) = \mu_0 + 1.086 (r/r_0)^n,
\end{equation}
with $\mu(r)$ the surface brightness at radius $r$ (we measure radii
as the geometric mean of the semi-major and semi-minor axes of the
isophotes, in arcsec), $\mu_0$ the extrapolated central surface
brightness (both in mag\,arcsec$^{-2}$) and $r_0$ the scale radius, in
arcsec. The S\'ersic shape parameter\footnote{Note that in some
studies the S\'ersic shape parameter is defined as the reciprocal
$1/n$.} $n$ quantifies the central concentration of the surface
brightness profiles, with $n=1/4$ corresponding to a centrally
concentrated de Vaucouleurs profile and $n=1$ to a diffuse
exponential. A $\chi^2$ fit to the profiles has been performed outside
4\,arcsec and above the level of 26\,mag\,arcsec$^{-2}$, to exclude the
emission clouds in the centre and to avoid the outer parts of the
galaxies which might be affected by flat-field and sky-subtraction
uncertainties. In order to compare our results to other studies which
use an exponential instead of a S\'ersic law, we also fitted an
exponential law to the same part of the surface profiles, deriving a
central surface brightness $\mu_0$ and a scale length $r_0$.  All
photometric parameters are listed in Table \ref{tab_photometry}.
\begin{table*}
  \begin{minipage}{168mm} 
    \begin{center} 
      \caption{Photometric parameters of FCC032, FCC206, and
	FCCB729}
      \label{tab_photometry}
      \begin{tabular}{lccccccccc}
\hline 
name & Type & $m_R$ & $r_\text{eff}$ & $\langle \mu \rangle_\text{eff}$ & $r_0$ (1) & $\mu_0$ (1) & $n$ (1) & $r_0$ (2) & $\mu_0$ (2) \\
 & & (mag) & (arcsec) & (mag\,arcsec$^{-2}$) & (arcsec) & (mag\,arcsec$^{-2}$) & &(arcsec) & (mag\,arcsec$^{-2}$) \\
\hline 
FCC032 & dE2 & 14.52 & 7.59 & 20.92 & 1.41 & 18.69 & 0.64 & 5.52 & 20.28 \\
FCC206 & dE0 & 15.02 & 15.86 & 23.01 & 15.33 & 22.43 & 1.33 &  10.64 & 22.02 \\
FCCB729 & S0(7),N & 15.15 & 7.21 & 21.44 & 5.19 & 20.71 & 1.04 &  4.84 & 20.62 \\
\hline 
      \end{tabular}
    \end{center}
    Galaxy type as classified by us, total de-reddened $R$-band
    magnitude $m_R$, half-light radius $r_{\rm eff}$, and
    effective surface brightness $\langle \mu \rangle_{\rm
      eff}$. (1):~parameters of the S\'ersic profile that best fits
    the surface brightness profile:~scale-length $r_0$,
    extrapolated central $R$-band surface brightness $\mu_0$, and
    shape-parameter $n$; (2):~parameters of the best fitting
    exponential profile:~scale-length $r_0$ and extrapolated
    central $R$-band surface brightness $\mu_0$.
  \end{minipage}
\end{table*}

\begin{figure}
  \includegraphics[width=84mm]{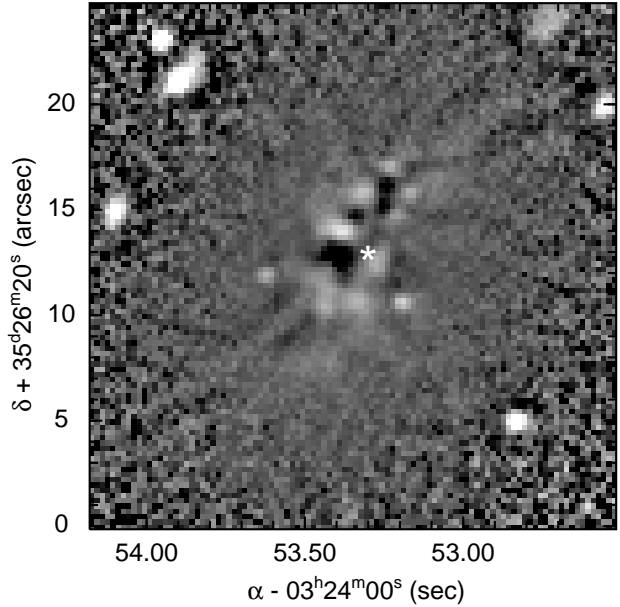}
  \caption{The $R$-band image of FCC032, divided by its median
    filtered homologue (running box of 1.25$\times$1.25\,arcsec).
    There are many bright spots (white), some of which are also
    visible in the H$\alpha+$[N\,\textsc{ii}] image (though not
    all). Some of these bright spots might be foreground or background
    objects (although such a strong concentration of chance alignments
    close to the galaxy centre seems highly unlikely) while others
    correspond to star formation sites that ionize the surrounding
    gas. Also clearly visible are the dark dust patches. The asterisk
    marks the centre of the outer isophotes.}
  \label{fig_032un}
\end{figure}
\begin{figure}
  \includegraphics[width=84mm]{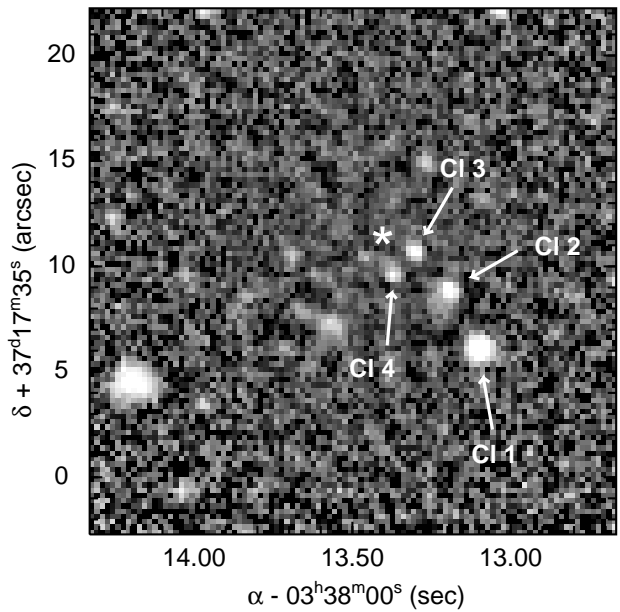}
  \caption{The $R$-band image of FCC206, divided by its median
    filtered homologue (running box of 1.25$\times$1.25\,arcsec). Four
    point sources or `star clusters' (named {\tt Cl1} to {\tt Cl4})
    can be discerned, two of which, {\tt Cl2} and {\tt Cl3}, also
    appear in the pure emission image in Fig. \ref{fig_206em}. There
    are no traces of dust absorption. The asterisk marks the centre of
    the outer isophotes.}
  \label{fig_206un}
\end{figure}
\begin{figure}
  \includegraphics[width=84mm]{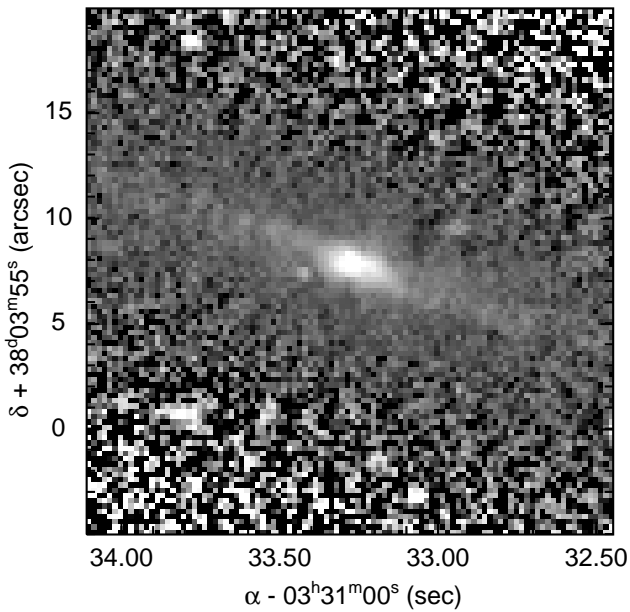}
  \caption{The $R$-band image of FCCB729, divided by its median
    filtered homologue (running box of 1.25$\times$1.25\,arcsec).  A
    faint stellar disc can be discerned which could also be due to the
    high degree of flattening of this dS0,N. The nucleus of
    FCCB729 appears slightly extended in the south-west direction,
    coincident with the extended H$\alpha$ emission
    (Fig. \ref{fig_729em}).}
  \label{fig_729un}
\end{figure}

\subsection{FCC032}
Based on blue-sensitive photographic plates taken with the Las
Campanas 2.5\,m du Pont telescope and the UK Schmidt Telescope,
\citet{ferguson89} classified FCC032 as dE~pec/BCD, most likely
inspired by the presence of several bright knots and dust patches in
the central region, which are also visible in our $R$-band images. We
applied an unsharp masking technique to the $R$-band images to highlight
possible small-scale structure \citep{derijcke03a}. The aforementioned
features stand out very clearly in the unsharp-masked image in
Fig. \ref{fig_032un}. None of these bright knots is at the centre of the
outer isophotes. We took the brightest knot (in $R$-band) as the
reference point, but the photometry in the inner 4\,arcsec is hampered
by these knots, as can be seen in the rapid variation of the Fourier
coefficients. FCC032's location in a cluster, its diffuse,
near-exponential surface brightness profile, and its regular,
elliptical isophotes argue in favour of a classification as a
non-nucleated dE2.

\subsection{FCC206}
\citet{caldwell87}, in their photometric study of a sample of Fornax
galaxies, noted several bright knots close to, but not actually at,
the centre of FCC206. We find four unresolved sources in the $R$-band
image. They show up very clearly in the unsharp masked image in
Fig. \ref{fig_206un}.  FCC206 is quite blue compared with other Fornax
dEs and its $B-V=0.52$ colour is redder than would be expected from
its $U-B=-0.16$ colour (assuming a single-age, single-metallicity
globular-cluster like population), indicative of a composite
population containing both cool stars and hot young stars. The knots
are found to be even bluer and with luminosities comparable to the
brightest and bluest Large Magellanic Cloud globular clusters. These
authors suggest that we are witnessing the formation of a
nucleus. \citet{caldwell87} estimate the absolute magnitudes of these
knots at $M_V \sim -8.5$. Assuming the clusters to reside inside
FCC206, we find $m_R({\tt Cl1})=21.7$, $M_R({\tt Cl1})=-9.7$;
$m_R({\tt Cl2})=23.7$, $M_R({\tt Cl2})=-7.6$; $m_R({\tt Cl3})=23.8$,
$M_R({\tt Cl3})=-7.5$; and $m_R({\tt Cl4})=23.7$, $M_R({\tt
Cl4})=-7.6$. The values of {\tt Cl2} and {\tt Cl3} are corrected for
the contribution of the H$\alpha$ emission to the $R$-band flux. ({\tt
Cl1} and {\tt Cl4} are absent in the H$\alpha+$[N\,\textsc{ii}]
images, see section \ref{sec_halpha}). The naming of the clusters is
clarified in Fig. \ref{fig_206un}. These absolute magnitudes are
similar to those of Galactic globular clusters, the brightest of which
is $\omega$Cen with $M_V=-10.6$. Due to its roundness, FCC206 has a
very ill defined PA, changing continuously over more than 240
degrees. Even outside the central region, where the photometry is not
affected by the bright knots, the position of the centre of the
isophotes varies significantly and rapidly. We constructed surface
brightness models keeping the centre coordinates fixed, but the
residuals are always larger compared to the model with varying centre
coordinates, thus we conclude the variation in $\Delta \delta$ and
$\Delta \alpha$ is real. In the following we take the centre of the
outer isophotes as the centre of the galaxy. Finally, FCC206 is
non-nucleated and has a very large flat core, which
is reflected in its large S\'ersic shape parameter $n = 1.33$.

\subsection{FCCB729}
Due to its relatively high surface brightness, FCCB729 was originally
classified as a background galaxy \citep{ferguson89}. However,
redshift measurements by \citet{drink01} identify FCCB729 as a true
Fornax Cluster member. Moreover, FCCB729 has a nearly exponential
surface brightness profile, typical of a genuine dwarf
galaxy. Subtracting the best fitting S\'ersic model off the surface
brightness profile, we can compute the luminosity of the nucleus. We
arrive at an apparent $R$-band magnitude $m_R=19.64$, or absolute
magnitude $M_R=-11.68$. The effect of seeing alone is not strong
enough to account for the decline in ellipticity towards the centre. A
simulated pure elliptical surface brightness distribution with
$\epsilon = 7$ convolved with an 0.7\,arcsec FWHM Gaussian point
spread function would still reach $\epsilon \simeq 5$ in the
centre. Thus the nucleus is significantly rounder than the galaxy
itself, although overall, FCCB729 is a very flattened system. Its
$C_4$ profile is systematically positive which indicates that the
isophotes are slightly disky, especially towards the outskirts.  This
argues for a classification as dS0(7),N, rather than dE7,N. The dip in
$C_4$ around $r \sim 17$\,arcsec is due to the two symmetrically
opposed bright knots along the major axis, which were masked to derive
the photometry. This results in a more boxy appearance at that point.
With the unsharp masking technique, we detected what seems to be a
faint stellar disc embedded in FCCB729 (Fig. \ref{fig_729un}),
although some caution is needed here. We also applied the technique to
the simulated pure dE7 and saw the same signature of a disc. This
signature is entirely due to the high degree of flattening and the
size of the filtering running box.

\section{H$\balpha$ imaging}  \label{sec_halpha}

\subsection{The H$\balpha$ equivalent width}

\citet{drink01} have measured H$\alpha$ EWs of 108 confirmed Fornax
Cluster members, including FCC032, FCC206 and FCCB729, with the
FLAIR-II spectrograph on the UK Schmidt Telescope. The effective
aperture of this system is at least 6.7\,arcsec (the fiber diameter)
and could be as large as 15\,arcsec because of image movements due to
tracking errors and differential atmospheric refraction. These authors
find:\\
\begin{tabular}{lcr}
  EW(FCC032)  & = &  9.6\,\AA,\\
  EW(FCC206)  & = & 14.4\,\AA,\\
  EW(FCCB729) & = &  6.8\,\AA.\\
\end{tabular}\\
For comparison, we calculated the EW inside an aperture radius $r$
from our images as :
\begin{equation}
  \text{EW} = \frac{F_\text{em}(r)}{F_\text{cont}(r)} \Delta \lambda,
\end{equation}
with $\Delta \lambda = 64$\,{\AA} the FWHM of the H$\alpha$ filter and
$F_\text{em}(r)$ and $F_\text{cont}(r)$ the total number of counts
inside a circular aperture with radius $r$ of the narrowband and the
continuum (scaled $R$-band) image, respectively. We find :\\
FCC032
\begin{equation*}
  \begin{array}{ll}
    F_\text{em}(3.5\,\text{arcsec})  & = 700 \,\text{e$^-$\,s$^{-1}$} \\
    F_\text{cont}(3.5\,\text{arcsec})& = 2340 \,\text{e$^-$\,s$^{-1}$}  \\
     & \rightarrow \text{EW}(3.5\,\text{arcsec}) = 19.1 \,\text{\AA} \\
    F_\text{em}(8\,\text{arcsec})  & =  968 \,\text{e$^-$\,s$^{-1}$}  \\
    F_\text{cont}(8\,\text{arcsec})& = 5920 \,\text{e$^-$\,s$^{-1}$}  \\
     & \rightarrow \text{EW}(8\,\text{arcsec}) = 10.5 \,\text{\AA}, \\
  \end{array}
\end{equation*}
FCC206
\begin{equation*}
  \begin{array}{ll}
    F_\text{em}(3.5\,\text{arcsec})  & = 100 \,\text{e$^-$\,s$^{-1}$} \\
    F_\text{cont}(3.5\,\text{arcsec})& = 355 \,\text{e$^-$\,s$^{-1}$} \\
     & \rightarrow \text{EW}(3.5\,\text{arcsec}) = 18.0 \,\text{\AA} \\
    F_\text{em}(8\,\text{arcsec})  & = 308 \,\text{e$^-$\,s$^{-1}$} \\
    F_\text{cont}(8\,\text{arcsec})& = 1300 \,\text{e$^-$\,s$^{-1}$} \\
     & \rightarrow \text{EW}(8\,\text{arcsec}) = 15.2 \,\text{\AA}, \\
  \end{array}
\end{equation*}
FCCB729
\begin{equation*}
  \begin{array}{ll}
    F_\text{em}(3.5\,\text{arcsec})  & = 280 \,\text{e$^-$\,s$^{-1}$} \\
    F_\text{cont}(3.5\,\text{arcsec})& = 1383 \,\text{e$^-$\,s$^{-1}$} \\
     & \rightarrow \text{EW}(3.5\,\text{arcsec}) = 13.0 \,\text{\AA} \\
    F_\text{em}(8\,\text{arcsec})  & = 302 \,\text{e$^-$\,s$^{-1}$} \\
    F_\text{cont}(8\,\text{arcsec})& = 3010 \,\text{e$^-$\,s$^{-1}$} \\
     & \rightarrow \text{EW}(8\,\text{arcsec}) = 6.4 \,\text{\AA}. \\
  \end{array}
\end{equation*}

In all, these values are in good agreement with the EWs measured by
\citet{drink01}, considering the uncertainties that affect both
measurements (photon shot-noise, sky and continuum subtraction,
positioning of the FLAIR-II spectrograph fiber on these faint
objects, \ldots).
\begin{figure}
  \includegraphics[width=84mm]{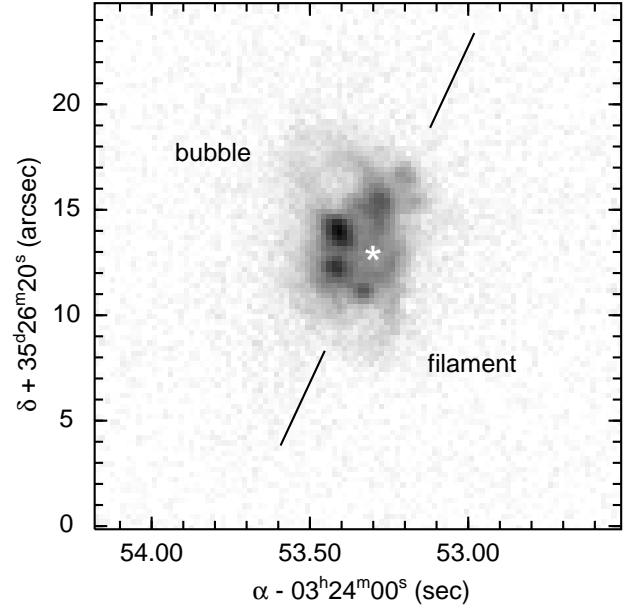}
  \caption{The pure H$\alpha$+[N\,\textsc{ii}] emission image of
    FCC032. Several emission clouds are visible, forming a gas complex
    elongated along the galaxy's major axis (the direction of the
    major axis is indicated by two black lines). Towards the
    north-east a superbubble is visible; towards the south, a gas
    filament (or bubble?), extending away from the galaxy centre, can
    be discerned. The asterisk marks the centre of the outer
    isophotes.}
  \label{fig_032em}
\end{figure}
\begin{figure}
  \includegraphics[width=84mm]{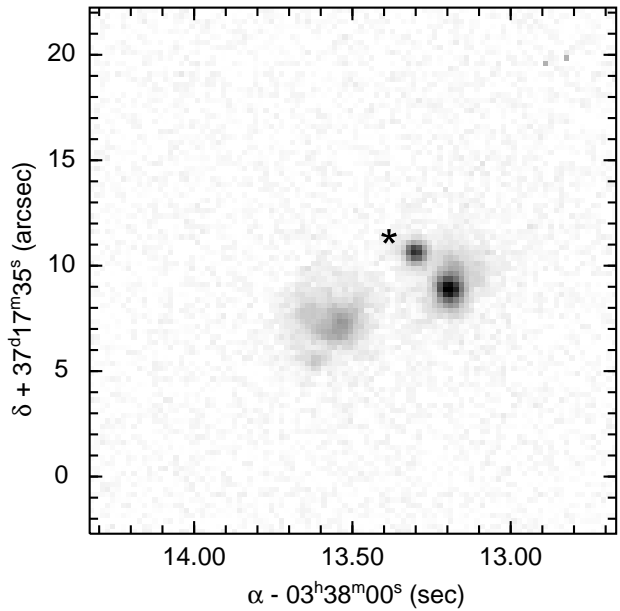}
  \caption{The pure H$\alpha$+[N\,\textsc{ii}] emission image of
    FCC206. Two bright emission clouds are visible and an extended
    emission region to the south-east. The asterisk marks the centre
    of the outer isophotes. The adopted grayscale is the same as in
    Fig. \ref{fig_032em}.}
  \label{fig_206em}
\end{figure}
\begin{figure}
  \includegraphics[width=84mm]{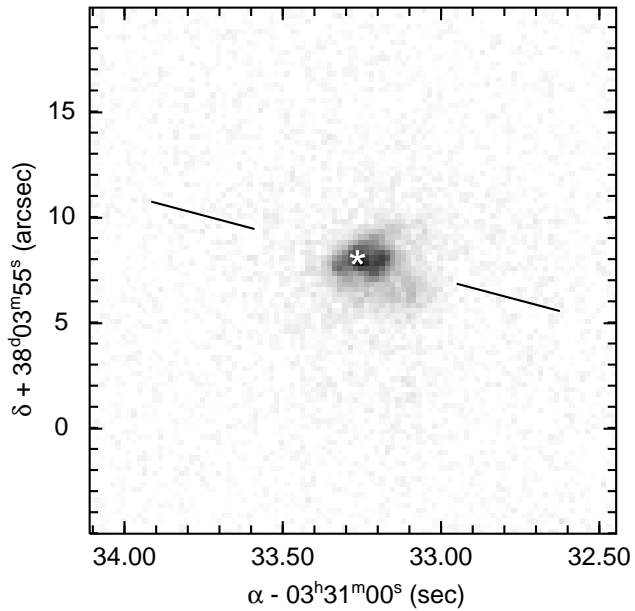}
  \caption{The pure H$\alpha$+[N\,\textsc{ii}] emission image of
    FCCB729. The central emission extends to the west, approximately
    along the direction of the major axis (indicated by two black
    lines). The asterisk marks the centre of the outer isophotes. The
    adopted grayscale is the same as in Figure \ref{fig_032em}.}
  \label{fig_729em}
\end{figure}

\subsection{The H$\balpha$+[N\,{\bf\sc ii}] and H$\balpha$ luminosities}

\begin{table*}
  \begin{minipage}{168mm}
    \begin{center}
      \caption{Emission properties of FCC032, FCC206, and FCCB729.}
      \label{tab_emission_overview}
      \begin{tabular}{lcccccc}
\hline
name & $F_\text{em}$ & $L_\text{em}$ & $F_{\text{H}\alpha}$ & $L_{\text{H}\alpha}$ & SFR & $M_\text{H\,\textsc{ii}}$ \\
 & $10^{-18}$\,W\,m$^{-2}$ & $h_{75}^{-2}\times 10^{31}$\,W & $10^{-18}$\,W\,m$^{-2}$ & $h_{75}^{-2}\times 10^{31}$\,W & $h_{75}^{-2}\times 10^{-3}$\,M$_\odot$\,yr$^{-1}$ & $h_{75}^{-2}$\,M$_\odot$ \\
\hline
FCC032  & 19.3 -- 22.0 & 7.81 -- 8.90 & 19.3 -- 6.00 & 2.43 -- 7.81 & 17.5 -- 5.5 & 570 -- 1850 \\
FCC206  & 6.17 -- 7.17 & 2.50 -- 2.90 & 1.96 -- 6.17 & 0.79 -- 2.50 & 1.8 -- 5.6 & 190 -- 590 \\
FCCB729 & 6.21 -- 7.33 & 2.51 -- 2.96 & 2.00 -- 6.21 & 0.81 -- 2.51 & 1.8 -- 5.6 & 190 -- 600 \\ 
\hline
      \end{tabular}
    \end{center}
  \end{minipage}
\end{table*}

\begin{figure}
  \includegraphics[clip,width=84mm]{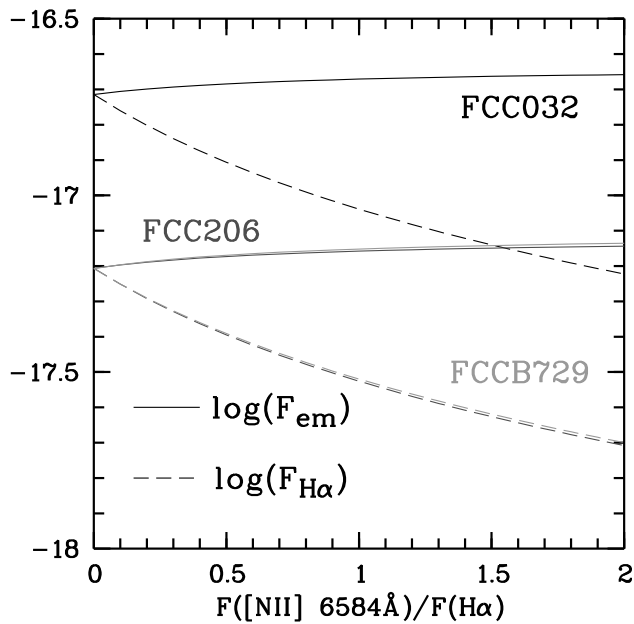}
  \caption{The logarithm of the total H$\alpha$+[N\,\textsc{ii}] flux
    $F_\text{em}$ (solid lines) and the H$\alpha$ flux
    $F_{\text{H}\alpha}$ (dashed lines) versus the ratio of the
    strengths of the [N\,\textsc{ii}] 6584{\AA} and the H$\alpha$
    line. The total flux is virtually independent of this
    line-ratio. The fluxes for FCC206 and FCCB729 are almost
    identical.}
  \label{fig_nitroverhalpha}
\end{figure}

\begin{figure}
  \includegraphics[clip,width=84mm]{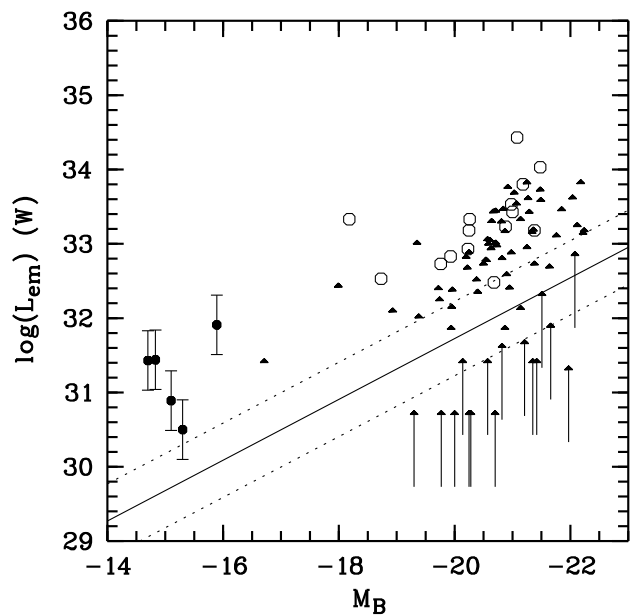}
  \caption{The total H$\alpha$+[N\,\textsc{ii}] emission-line
  luminosity versus absolute blue magnitude. The black dots are our
  sample of dEs (this paper and Paper~1); the circles and triangles
  represent the Es and S0s observed by \citet{buson93} and
  \citet{macchetto96}, respectively. The lines indicate the linear
  relation and its $1-\sigma$ relation observed by
  \citet{phillips86}. All observations have been corrected to the
  distance scale adopted in this paper.}
  \label{fig_buson}
\end{figure}

We present the pure H$\alpha$+[N\,\textsc{ii}] images of FCC032,
FCC206 and FCCB729 in Figs. \ref{fig_032em}, \ref{fig_206em} and
\ref{fig_729em}. The emission fluxes are calculated using
eqs. \eqref{eq_fluxha} and \eqref{eq_fluxem} and depend on the adopted
value for $F_{\text{[N\,\textsc{ii}]}_2}/F_{\text{H}\alpha}$. In
Fig. \ref{fig_nitroverhalpha} we show how the total emission and the
H$\alpha$ fluxes vary with
$F_{\text{[N\,\textsc{ii}]}_2}/F_{\text{H}\alpha}$ going from 0 to
2. The H$\balpha$+[N\,{\bf\sc ii}] and H$\balpha$ fluxes and
luminosities are listed in columns 2 -- 5 of Table
\ref{tab_emission_overview}. In figure \ref{fig_buson}, we compare the
total emission luminosity with those of FCC046 and FCC207 (Paper~1)
and with those found for elliptical and S0 galaxies by
\citet{phillips86}, \citet{buson93} and \citet{macchetto96}. The
linear correlation between emission-line luminosity $L_\text{em}$ and
absolute blue magnitude $M_B$ in the large sample of galaxies studied
by \citet{phillips86} strongly suggests that the dominant ionising
source in 'normal' early-type galaxies is a component of the stellar
population. In these galaxies, the emission-line region is mostly
concentrated in the centre, while galaxies with extended emission
\citep{buson93} have emission-line luminosities 10 -- 100 times larger
than 'normal' Es and S0s.  This increase in emission luminosity
requires a second ionising source to be present. This can be
photo-ionisation by young hot stars or shock-ionisation by supernova
remnants in those galaxies in which recent star-formation is
present. On the other hand, an active nucleus can also provide an
extra source of ionisation. The dEs in our sample extend the
correlation of $L_\text{em}$ versus $M_B$ for early-type galaxies with
extended emission.

In FCC032, the emission is indeed extended and comprises several emission
clouds. The total flux of FCC206 is about 3 times lower. In this
galaxy, three separate emission clouds are clearly visible. Finally,
the emission of FCCB729 is concentrated in the centre, with a flux
comparable to that of FCC206. This is the only galaxy in which the
emission coincides with the stellar nucleus; the other two show
distinct emission clouds {\em around} the centre. \citet{binette94}
propose photo-ionisation by post-AGB stars in an old stellar
population as a source for the central emission in elliptical
galaxies. Using their prescriptions, we derive central H$\alpha$
luminosities of the order of $2 \times 10^{30}$\,W, i.e. a factor
$5-10$ less than what is observed here, suggesting that other
ionisation mechanisms are (also) present. If, on the other hand, all
H$\alpha$ emission were due to photo-ionisation by young stars, we can
estimate the star formation rate (SFR) using the calibration between
the total SFR and the H$\alpha$ luminosity derived by
\citet{kennicutt83}:
\begin{equation}
  \text{SFR} \simeq 8.93 \times 10^{-35} L_{\text{H}\alpha} E_{\text{H}\alpha}\,\text{M}_\odot\,\text{yr}^{-1},
\end{equation}
where $E_{\text{H}\alpha} = 2.512$ is the standard 1~magnitude factor
to correct for internal extinction and $L_{\text{H}\alpha}$ is the
H$\alpha$ luminosity expressed in W. The obtained SFR varies between
$\sim 10^{-3} - 10^{-2}$\,M$_\odot$\,yr$^{-1}$ in these galaxies (see
Table \ref{tab_emission_overview}, column 6).


\begin{figure}
  \includegraphics[width=84mm]{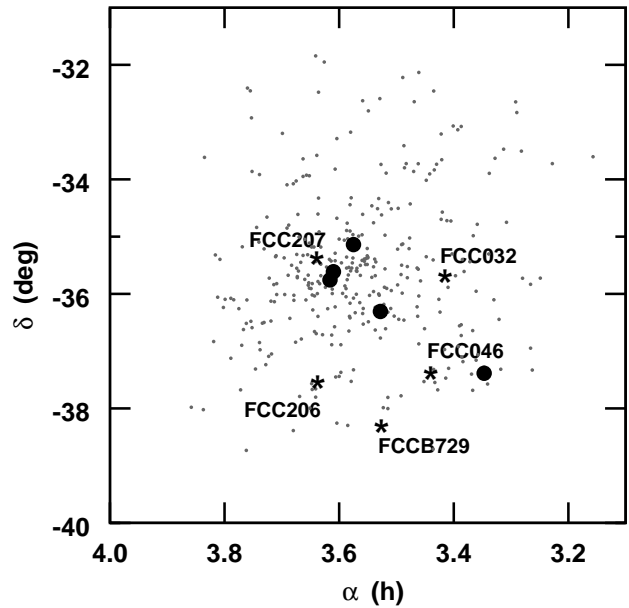}
  \caption{Positions of the five dEs containing ionized gas imaged by
    us so far within the Fornax cluster. Small dots indicate the
    positions of 340 galaxies in the Fornax Cluster Catalogue. Large
    dots indicate the positions of galaxies brighter than $M_B <
    -20$. Asterisks indicate the positions of FCC032, FCC046, FCC206,
    FCC207, and FCCB729. Except for FCC207, which has a projected
    position close the cluster centre, these dEs appear to populate
    the less densely populated outskirts of the cluster.} 
  \label{fig_cat}
\end{figure}

The H$\alpha+$[N\,\textsc{ii}] luminosities can also be compared
with those observed in the Local Group dEs:~NGC147, NGC185, and
NGC205, three companions of M31. \citet{yl97} have detected a central
emission region and several compact emission objects in NGC185 with a
total H$\alpha+$[N\,\textsc{ii}] luminosity $L_\text{em}=1.3 \times
10^{29}$\,W. This is still an order of magnitude smaller than the
H$\alpha+$[N\,\textsc{ii}] luminosities of the dEs presented here. No
extended H$\alpha$ emission was detected in NGC205. NGC185 and NGC205
both contain a cold interstellar medium while NGC147 appears to be
devoid of gas \citep{sage98}. Hence, the Fornax dEs presented in
Paper~1 and in this paper, are comparatively rich in
ionized gas. The stark contrast in environment (these dEs reside
mostly in the outskirts of the Fornax cluster, see Fig. \ref{fig_cat},
while the M31 satellites are members of a rather compact group,
orbiting a massive spiral galaxy) suggests that external factors, such
as interactions, affect the gas-depletion rates of dwarf
galaxies. N-body/SPH models of dwarf galaxies orbiting inside the
Milky Way halo \citep{mayer01} show that their gas reservoirs are
depleted within a few gigayears after the first pericentre passage,
due to both tidal stripping and star bursts occurring after each
pericentre passage. These results for the Milky Way and its satellites
should a fortiori be valid for M31 and its companions, given that they
have been on bound orbits long enough.

\subsection{H\,{\bf\sc ii} masses}

The total mass in ionized hydrogen, $M_\text{H\,\textsc{ii}}$, can be
written as
\begin{equation}
  \label{eq_hmass}
  M_\text{H\,\textsc{ii}} = 
  \frac{ L_{\text{H}\alpha} }{ 4\pi  j_{\text{H}\alpha} } m_\text{H} N_e ,
\end{equation}
with $L_{\text{H}\alpha}$ the total H$\alpha$ luminosity, $m_\text{H}$
the mass of the hydrogen atom and $N_e$ the electron density in the
gas. The hydrogen H$\alpha$ emissivity $j_{\text{H}\alpha}$ is given
by \citep{osterbrock89}
\begin{equation}
  \label{eq_emissivity}
  4\pi j_{\text{H}\alpha} = N_e^2 \alpha_{\text{H}\alpha}
  h\nu_{\text{H}\alpha} = 3.544 \times 10^{-32} N_e^2 \,\text{W\,cm}^{-3}
\end{equation}
in 'case B' recombination, i.e. complete re-absorption of all Lyman
photons in an optically thick nebula. Each Lyman photon emitted from a
level with $n \geq 3$ is later on converted to (a) Balmer photon(s)
plus one Lyman $\alpha$ photon, thus raising the flux in the Balmer
lines. The production coefficient $\alpha_{\text{H}\alpha}$
(calculated for $T = 10^4$\,K) is insensitive to the electron density
(it changes by only 4 per cent if $N_e$ is raised from 1\,cm$^{-3}$ to
$10^6$\,cm$^{-3}$) and varies as $T^{-0.8}$ as a function of
temperature. Using equations \eqref{eq_hmass} and
\eqref{eq_emissivity}, the ionized hydrogen mass can be written
concisely as \citep{kim89}:
\begin{equation}
  \label{eq_hmassbis}
  \begin{array}{lcl}
    M_\text{H\,\textsc{ii}} & = & 23.72 
      \left( \frac{ 1000\, \text{cm}^{-3} }{N_e} \right) 
      \left( \frac{ L_{\text{H}\alpha} }{10^{30}\,\text{W}} \right)
      \, \text{M}_\odot \\
      & = &  2.85 
      \left( \frac{ 1000\, \text{cm}^{-3} }{N_e} \right)
      \left( \frac{ F_{\text{H}\alpha} }{10^{-19}\, \text{W\,m}^{-2}} \right) 
      \left( \frac{D}{10\, \text{Mpc}} \right)^2 \text{M}_\odot ,
  \end{array}
\end{equation}
with $D$ the distance to the galaxy.

In the following, we will assume the value $N_e = 1000$\,cm$^{-3}$ for
the electron density to be in accord with most other authors and to be
able to directly compare our ionized hydrogen masses with the
literature (however, \citet{spitzer78} advocates $N_e =
100$\,cm$^{-3}$ as a typical value for both Galactic H\,\textsc{ii}
regions with diameters of the order of 100\,pc and for supernova
remnants). Using equation \eqref{eq_hmassbis}, we derived ionized
hydrogen gas masses between $10^2$ and $10^3$\,M$_\odot$ (see column 7
of Table \ref{tab_emission_overview}).

\section{Discussion} \label{sec_disc}

\subsection{FCC032}

FCC032 is the most gas-rich galaxy in our sample and contains an
extended ionized gas complex, about 10\,arcsec ($\approx 850$\,pc)
across. The bulk of the H$\alpha$ emission in FCC032 is distributed
over 4 different emission clouds, 2 of which are clearly visible in
the $R$-band. The other 2 are only visible in the pure-emission
images. These clouds appear to lie on a semi-ellipse around the centre
of the outer isophotes, with its major axis along the major axis of
the galaxy. To the north of the centre, a superbubble is visible, with
a diameter of 350\,pc, capped with a $\sim 90$\,pc thick
shell. Towards the south, a gaseous filament (or bubble?)  is
extending $\sim 260$\,pc away from the galaxy centre.  A comparison of
Fig. \ref{fig_032em} with Fig. \ref{fig_032un} shows that the
low-surface brightness cavities in Fig. \ref{fig_032em} cannot be due
to dust-absorption. This image bears a very strong resemblance to
Fig. 1, panel b, in \citet{calzetti04}, which shows a high-resolution
HST narrow-band image of the H$\alpha$ emission in the starbursting
dwarf galaxy NGC3077. The diameter of the bubble in FCC032 is
comparable to those found in hydrodynamical simulations of supernova
remnants (SNRs) \citep{db01} and those observed in irregular dwarf
galaxies by \citet{martin98}, who measures (super)bubble diameters of
$\sim 100$ to 2000\,pc.

Such shells of ionized gas are quite common in more gas-rich and
star-forming dwarf galaxies such as dwarf irregulars. Usually, these
shells consist of gas that is shock-ionized by the supernova
explosions and stellar winds leaving the starburst regions. In all,
there appears to be a very strong energy feedback of the starburst
into the interstellar medium. The strong resemblance between the
shells in starbursting dwarf galaxies such as NGC3077 and the ones
observed in FCC032, are strong evidence for recent or ongoing
star-formation in this dE.

\subsubsection{H\,\textsc{i} observations}

We also observed FCC032 with the Australia Telescope Compact Array
(ATCA) in May 2003 for a total of $\approx4$\,hours.  The array
configuration used was the EW352, and an 8 MHz-wide band was centered
upon 1413 MHz. The number of channels used was 512, giving a frequency
resolution of 15.6\,kHz, equivalent to 3.3\,km\,s$^{-1}$ per channel.
Sources 1934-638 and 0438-436 were also observed, the former for flux
and bandpass calibration, and the latter for phase and gain
calibration. All data reduction was carried out using MIRIAD.

We detected no H\,\textsc{i} gas in FCC032 and derived an upper
H\,\textsc{i} mass limit for the galaxy using:
\begin{equation}
  M_{\text{H\,\textsc{i}}} = 
  2.356 \times 10^{5} D^{2} \int{S(v)\,dv}\,\text{M}_\odot,
\end{equation}
where $M_{\text{H\,\textsc{i}}}$ is the H\,\textsc{i} mass, $D$ is the
distance to the galaxy in Mpc, and $\int{S(v)\,dv}$ is the integrated
flux in Jy\,km\,s$^{-1}$. Assuming a velocity width of the
galaxy of 50\,km\,s$^{-1}$, a 3$\sigma$ detection would have an upper
H\,\textsc{i} mass limit of $7.1\times10^{7}$\,M$_\odot$.

In an H\,\textsc{i} study of 6 BCDs with a dE-like appearance,
\citet{vanzee01} found H\,\textsc{i} masses ranging between $0.35 -
4.23 \times 10^{8}$\,M$_{\odot}$. \citet{conselice03} detected
H\,\textsc{i} in 2 Virgo dEs with H\,\textsc{i} masses of 6 and 8$
\times 10^{7}$\,M$_{\odot}$ which is around our detection limit. The
non-detection of H\,\textsc{i} in FCC032 again argues for its
classification as dE, rather than BCD.

\subsection{FCC206}

The emission in FCC206 comes from 3 separated emission clouds. The two
brightest clouds are unresolved. There is also an extended emission
region, about 4\,arcsec ($\approx 350$\,pc) across, to the south-east
of the centre. {\tt Cl2} and {\tt Cl3} are the only point sources in
the $R$-band image that coincide with an emission region. {\tt Cl1}
and {\tt Cl4} could either be foreground stars or young globular
clusters inside FCC206. Their blue colours, similar to the bluest
Large Magellanic Cloud globular clusters, as reported by
\citet{caldwell87}, seem to argue for the latter interpretation
although, clearly, only via spectroscopy can one shed light on the
true nature of these point sources.

\subsection{FCCB729}

FCCB729 has an extended emission region in the centre, with faint
extensions towards the west, along the galaxy's major axis, and
towards the north, along its minor axis. The emission peak coincides
with the galaxy's central nucleus.

\section{Conclusions}
\label{sec_concl}

\subsection{Ionisation mechanisms}

Combining the results of this paper and Paper~1, we find different
ionized gas morphologies in these galaxies. This morphological
diversity could also indicate a diversity in ionising processes.

All nucleated dEs in our sample, i.e. FCC046, FCC207 (Paper~1), and
FCCB729 have an emission peak coinciding with their central nucleus,
while the non-nucleated dEs lack such a bright central emission,
suggesting a physical connection between the central emission and the
presence of a nucleus. We are planning follow-up spectroscopy for
these objects in order to identify the ionisation mechanism in the
central regions (active galactic nucleus (AGN), starburst, post-AGB
stars,\ldots). The discovery of an AGN in a bona-fide dE would be of
considerable interest for theories of the formation and evolution of
dEs and their nuclei. Up to now, only one AGN in a dE has been
reported:~\citet{barth04} show evidence for a Seyfert~1 nucleus in the
bright dE/faint E POX~52, its classification as either a dE or E made
uncertain by its low S\'ersic shape parameter $n=0.28$, which means
the galaxy is well described by a de Vaucouleurs profile, and its high
luminosity. On the other hand, its position in the Fundamental Plane
argues for a dE classification.

In three dEs of our sample (FCC032, FCC046 and FCC206) the emission
comprises several emission clouds that argue for recent or ongoing
star formation (H\,\textsc{ii} clouds, SNRs,\ldots). In particular
the ionized gas complex in FCC032 is similar to those observed in
star-forming dwarf irregulars. Such dEs could be
descendants of more fiercely star-forming dwarf galaxies, such as Blue
Compact Dwarfs, which are not (or no longer) present in the Fornax
Cluster. The fact that the number density of dEs with H$\alpha$ or
H\,\textsc{i} emission declines as a function of radius within their
host cluster \citep{drink01,conselice03}, indicates that environmental
agents, such as ram-pressure stripping or gravitational interactions,
play a very important role in the gas-depletion process.

\subsection{Formation of the central nucleus}

FCC206 is a non-nucleated dE but it contains 4 non-resolved star
clusters, two of which show H$\alpha$ emission. These knots have blue
colours and luminosities comparable to globular clusters, which tempted
\citet{caldwell87} to suggest we are witnessing the formation of a
nucleus. This corroborates one currently popular hypothesis concerning
the formation of nuclei in dEs, namely the merger of globular
clusters that have been driven inward by dynamical friction
(\citet{ol00}, \citet{lotz01}). \citet{mt03} have estimated the
orbital decay timescale of a typical globular cluster of mass $M \sim
10^6 $\,M$_\odot$, starting at a radius of about 1\,kpc in a dwarf
galaxy with a circular velocity of order $50-100$\,km\,s$^{-1}$ at
$1-5 \times 10^9$\,yr. The tidal disruption timescale is much longer
so a globular cluster would not dissolve before reaching the galaxy
centre where it adds its stars to the nucleus. If the point sources
near the centre of FCC206 are truly globular clusters belonging to
this dE, their projected distances of the order of 100\,pc would imply
the formation of a nucleus within the next $10^8$\,yr. Nucleated dEs,
after being stripped from their stellar envelope by tidal forces
during gravitational interactions, have been suggested as possible
progenitors of both the recently discovered ultra compact dwarfs
\citep{ph01} and of massive globular clusters like $\omega$Cen
\citep{gnedin02} and G1 in M31 \citep{meylan01}.

Intermediate-mass black holes (IMBHs), with masses of the order
$M_\bullet \sim 10^3 $\,M$_\odot$, are predicted to grow in dense star
clusters \citep{pzm02}. According to this scenario, a massive black
hole will grow in the nucleus of the host dE by the coalescence of
several globular cluster IMBHs. If globular clusters contain massive
black holes (e.g. \citet{gebhardt02}, \citet{gerssen02}), which is
still debated, and the nuclei in dE,Ns form from merging globular
clusters, these nuclei could be expected to contain super-massive
black holes (SMBHs). Thus understanding how the nuclei of dE,Ns form
may also help us understand how SMBHs grow.

\section*{Acknowledgments}
This work is based on observations collected at the European Southern
Observatory, Chile (ESO Programme Nr. 072.B-0134). We would like to
thank the service mode observers at Paranal for the excellent data. DM
acknowledges DM acknowledges the financial support of the Bijzonder
OnderzoeksFonds (BOF, Ghent University). SDR wishes to thank
V. Debattista, A. Pasquali, and I. Ferreras for fruitful discussions
about the origin and evolution of dE,N nuclei. WWZ acknowledges the
support of the Austrian Science Fund (project P14753). This research
has made use of the NASA/IPAC Extragalactic Database (NED) which is
operated by the Jet Propulsion Laboratory, California Institute of
Technology, under contract with the National Aeronautics and Space
Administration.

\label{lastpage}

\end{document}